\documentclass[pre,twocolumn, showpacs]{revtex4}

\usepackage{hyperref,graphicx}
\usepackage{amssymb}
\usepackage{latexsym}

\newcommand{\dv}{{\rm div\,}}
\newcommand{\f}{\vec{F}}
\newcommand{\ff}{F^2}
\newcommand{\fpa}{F_{\parallel}}

\newcommand{\fpp}{F_{\perp}^2}
\newcommand{\x}{\vec{x}}
\newcommand{\h}{\vec{h}}
\begin{document}

\title{\bf Activated sampling in complex materials at finite temperature: the
  properly-obeying-probability activation-relaxation technique}

\author{Henk Vocks$^2$, M.V. Chubynsky$^1$, G.T. Barkema$^2$ and
  Normand Mousseau$^1$\footnote{Permanent address}$^2$}

\affiliation{
        $^1$D\'epartement de Physique,
	Universit\'e de Montr\'eal,
        C.P. 6128, Succursale Centre-ville Montr\'eal,
        Qu\'ebec, Canada H3C 3J7}
\affiliation{ 
	$^2$Institute for Theoretical Physics, 
        Utrecht University, Leuvenlaan 4, 3584 CE Utrecht, 
        the Netherlands}
	
\date{\today}

\begin{abstract}
  While the dynamics of many complex systems is dominated by activated events,
  there are very few simulation methods that take advantage of this fact.  Most
  of these procedures are restricted to relatively simple systems or, as with the
  activation-relaxation technique (ART), sample the conformation space
  efficiently at the cost of a correct thermodynamical description. We present
  here an extension of ART, the properly-obeying-probability ART (POP-ART), that
  obeys detailed balance and samples correctly the thermodynamic ensemble. Testing POP-ART on
  two model systems, a vacancy and an interstitial in crystalline silicon, 
  we show that this method recovers the proper thermodynamical
  weights associated with the various accessible states and is significantly
  faster than MD in the diffusion of a vacancy below 700 K.
\end{abstract}


\maketitle

\section{Introduction}

Throughout physics, chemistry and biology, a large proportion of atomistic
processes take place on time scales many orders of magnitude longer than the
typical period of atomic vibrations.  These processes are out of reach of
straightforward molecular dynamics (MD), which cannot exceed simulation times
equivalent to the microseconds at best.  It is not surprizing, therefore, that
considerable effort has been devoted in the last few years, with some degree of
success, to develop algorithms that overcome this 
limitation~\cite{barkema96,doye97,voter97,voter97b,voter98,wu98,munro99,henkelman99,malek00,laio02}.

These methods can be separated into two classes. Activated methods, such as the
activation-relaxation technique (ART)~\cite{barkema96,malek00} and related
approaches~\cite{doye97,munro99,henkelman99} sample the energy landscape of
complex systems by identifying minima connected by minimum energy pathways. 
This family of methods is very efficient for sampling conformations.
Recently, ART was found to be the most efficient method for high-dimensional
problems~\cite{olsen04}. However, because of an unknown bias in the selection of events
for these methods, it is not possible to ensure a proper thermodynamic
sampling. While this is not a major limitation for sampling states or even
identifying pathways, such as protein folding trajectories, for example, it is
sufficiently severe to prevent the use of these methods to sample equilibrium or
dynamical quantities.

The second class of methods is based on molecular dynamics and corresponds to
methods such as hyper-MD~\cite{voter97}, temperature-assisted dynamics
(TAD)~\cite{voter97b}, self-guided dynamics~\cite{wu98} and biased
thermodynamics~\cite{laio02}. Until now, the application of these methods has
been mostly restricted to simple systems, with a limited number of relatively
well-characterized barriers or with a well-defined reaction
coordinate. Recently, Choudhary and Clancy have proposed modifications
of hyper-MD that could allow its application to disordered
materials~\cite{choudhary05}. It appears, however, that applying a significant
boost in hyper-MD could break the thermodynamical character of the algorithm,
placing this method in the first category of activated methods.

In this paper, we present an algorithm that offers correct thermodynamical
sampling without suffering from the usual exponential slowing down with
decreasing temperature. The properly-obeying-probability ART (POP-ART) samples
the thermodynamically relevant parts of phase space, hopping efficiently over
high barriers separating low-energy basins. We apply this algorithm to two test
cases, the diffusion of a self-interstitial and the self-diffusion of a vacancy
in Stillinger-Weber silicon~\cite{stillinger85}, to verify the correctness of 
the method, and to assess its efficiency.

The paper is organized as follows. We start with a brief discussion of
limitations of standard activated methods, such as ART. We then introduce
POP-ART and show how it can overcome these limitations and ensure a proper
thermodynamical sampling. The justification for the various steps needed to
construct activated pathways with detailed balance is then discussed in detail
and a summary of the algorithm is given.  The algorithm is tested in a study of
the diffusion of an interstitial and a vacancy in Stillinger-Weber silicon. In
Appendix A, we discuss a physical interpretation of the Jacobian used in
POP-ART; in Appendix B, we present an analytical calculation for a simple model
potential that provides further insights into the method;

\section{Sampling the energy landscape using activated methods}

The energy landscape of a system of $M$ atoms can be represented as an
$N\equiv 3M$-dimensional hypersurface, with the ``height'' indicating the value of the
potential of the configuration at a given set of atomic coordinates. In a
dynamical regime dominated by rare events, a system spends most of its time
oscillating near a local energy minimum, hopping over an energy barrier only
when a thermal fluctuation transfers sufficient energy onto the corresponding
mode. Since the probability of energy transfer decreases exponentially with its size,
the activated trajectory will tend to cross near the lowest-energy point on the
ridge, corresponding to a first-order saddle point.

It is possible to reconstruct these trajectories, as a sequence of local minima
separated by transition points, using the activation-relaxation
technique~\cite{barkema96} or related
methods~\cite{doye97,munro99,henkelman99}. In its latest form, called ART
nouveau~\cite{malek00}, this method works in three steps: (1) Starting from a
local minimum, a deformation is applied until the lowest curvature of the
Hessian matrix, given by
\begin{equation}
H_{ij} = \frac{\partial^2 E}{\partial x_i \partial x_j},
\end{equation}
becomes negative. (2) The configuration is then pushed along the corresponding
direction while the energy in the perpendicular hyperplane is minimized until it
reaches a first-order saddle point. (3) The configuration is then pushed
slightly further, away from this saddle point, and its energy is minimized using
a standard minimization technique.

ART and similar methods have been applied with success to study the topology of
the energy landscape and activated mechanisms in a wide range of materials
including amorphous and crystalline
semiconductors~\cite{barkema98,elmellouhi03,middleton01}, glassy
materials~\cite{mousseau00a}, atomic clusters~\cite{doye99a,malek00}, and
proteins~\cite{malek01,mortenson01,wei02,wei04a,melquiond05}. A recent study has
shown that ART compares favorably with other related techniques in terms of
efficiency and completeness of finding activated mechanisms~\cite{olsen04}.

It is tempting to use ART approaches to study dynamical trajectories, generating
an on-the-fly catalog of events that can be selected using kinetic Monte Carlo.
This approach was used by Henkelman and J\'onsson in the simulation of growth on
an Al (100) surface~\cite{henkelman01}, and by Middleton and
Wales~\cite{middleton04} for a binary Lennard Jones glass.  Comparing with
molecular dynamics, Middleton and Wales observed that the kMC results with an
on-the-fly catalog are qualitatively incorrect. It is likely that this
discrepancy is at least partly due to an uncontrolled selection bias for these
types of methods, making it impossible to ensure a proper sampling of the
barriers generating the catalog of kMC events.

These limitations are fundamental and cannot be lifted by simply doubling or
tripling the sampling. It is essential to incorporate thermodynamics at the core
of the algorithm. One method for achieving this is discussed in the next
section.

\section{The POP-ART approach}

We start by separating the configuration space into two regions: the {\it
basin} and the {\it saddle regions}. Basins are defined in a way that ensures
that they contain most of the thermodynamical weight at a
given temperature, while the saddle regions are visited only rarely and in
passing. The dividing (hyper-)surface between these two regions is chosen to be
the surface where the lowest curvature of the potential energy surface
equals a threshold value $\lambda_0$.  The basins represent the parts of the
configuration space where the lowest curvature has a value {\it above}
$\lambda_0$; they form a series of disconnected regions surrounding local
minima. The saddle region is on the other side of the threshold and includes
most other local extrema, such as first- and higher-order saddle points (see
Fig.~\ref{fig:sketch}).  The use of this criterion for separating the
configuration space is convenient as the status of any point in the
configuration space can be decided
locally, without relaxing to a nearby stationary point. For a given
threshold, it is always possible that the negative eigenvalue associated with a
particular saddle is higher (lower by absolute value) than the chosen threshold,
such that it belongs to a basin. As will be seen from the treatment of the
method below, this does not invalidate the algorithm but may even be used to
one's advantage.

Having separated the energy landscape, we define motion in each of these
regions. All the motion within the basin is performed with conventional
MD at the desired temperature. Once the configuration hits the dividing
surface, the MD is halted, the configuration is brought through the saddle
region to a new basin at the same energy, according to the activation
rules described below, and the MD is resumed at the same temperature. All
steps respect detailed balance and the overall trajectory samples
the basins according to the proper thermodynamical ensemble.

The activated part of the algorithm is composed of two steps: (1) the activation
trajectory is first generated, from one basin to the other, and then (2)
the free energy difference between the beginning and the end of this
trajectory is calculated. The latter information is used for the accept-reject step.

In the next subsections, we discuss these two steps in detail before presenting
the algorithm as it is currently implemented.

\subsection{The activated trajectory}

As in ART, an activation trajectory is created by moving along the eigenvector
corresponding to the lowest eigenvalue of the Hessian. Unlike ART, however,
there is no relaxation in the perpendicular hyperplane. Instead, all atoms are
moved in such a way as to keep the total potential energy constant. This is
easily achieved as the configuration is thermalized, with roughly $k_BT/2$
of available potential energy per degree of freedom above that in the local
energy minimum.  More specifically, the activated trajectory is generated by
iterating the following equation:
\begin{equation}
  \x_{i+1} = \x_i + \frac{\Delta \tau}{2}
  \left(\h_i+\h_{i+1}\right) + \frac{c\Delta \tau}{2}
  \left(\f_i+\f_{i+1}\right),
\label{eq:follow}
\end{equation}
where, $\h_i$ is the normalized eigenvector at $\x_i$ corresponding to
the lowest Hessian eigenvalue, $\f_i$ is the total
$N$-dimensional force at $\x_i$, $\Delta \tau$ is a constant
that determines the size of the increment, and $c$ is a multiplicative
constant, chosen to project the trajectory onto the hypersurface of
constant potential energy.  

The orientation of $\h_0$ is chosen initially so as to point in the direction of
more negative curvature, i.e., away from the initial basin; it is updated at
each step by requiring that the inner product of the local eigenvector $\h_i$
with that at the previous step, $\h_{i-1}$, be always positive. Values of $\h$
and $\f$ at point $\x_{i+1}$ are obtained iteratively, i.e., initially
$\h_{i+1}=\h_i$ and $\f_{i+1}=\f_i$ are used in Eq.~(\ref{eq:follow}) to obtain
a value of $\x_{i+1}$, then values of $\h$ and $\f$ are calculated at the new
point and inserted into Eq.~(\ref{eq:follow}) to get the next iteration, etc.

\begin{figure}
\includegraphics[width=8.5cm]{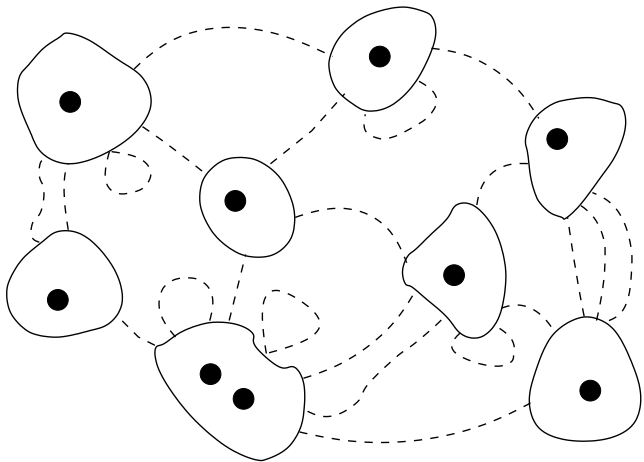}

\caption{Sketch of a two-dimensional energy landscape.  The black dots denote
  the locations of local-energy minima. These minima are part of basins, bounded
  by a line of constant lowest-curvature (solid line); the percolating region
  surrounding the basins is called the saddle region.  Basin-to-basin
  trajectories as generated by THWART are indicated by dashed lines.
  Constrained to ensure detailed balance, the trajectories come back to were
  they started if they fail to find a boundary. }
\label{fig:sketch}
\end{figure}

Unlike in ART, there is no separate relaxation stage, and Eq.~(\ref{eq:follow})
is iterated across the saddle region until the new basin is reached, i.e., until
the lowest eigenvalue passes the threshold (from below, this time). At this
point, the activation-relaxation phase is stopped and MD is resumed starting
with the new configuration (see Fig.~\ref{fig:sketch}).  This ensures that the
path generated from $\x_0$ to $\x_p$ is fully reversible: a configuration in
basin $p$ reaching $\x_p$ would trigger the activation, bringing it to the other
end of this path, at $\x_0$.  Reversibility, in a weak sense, is ensured by the
symmetric criterion for entering and leaving the saddle region as well as by
keeping the path on a hypersurface of constant energy: for each transition, its
inverse is also possible. In addition to reversibility, we have to ensure that
the relative probabilities of these transitions are correctly weighted; this is
discussed below.

The conservation of energy requires that the length of the velocity vector as MD
is restarted is equal to that at the beginning of the activation
($|\vec{v}_p|=|\vec{v}_0|$); the direction should be chosen so as to point
inside the new basin, but otherwise is arbitrary.  After entering the new basin,
MD is continued for a very small number of steps --- 10 or so --- to prevent the
system from a quick recrossing back to the original basin. This is implemented
by letting the system bounce back against the constant eigenvalue surface.
After these few steps, the MD stage continues until the system crosses the basin
boundary, and another activation is begun, etc.

We note that the activation path does not always lead to a new basin.  In our
simulations, as is illustrated in Fig.~\ref{fig:sketch}, it is not rare to see the
trajectory form a circular path, coming back very close to the initial point
$\x_0$, after a long excursion in the saddle region. It is also possible that
the trajectory returns to the same basin but not at the starting point. This
does not invalidate the algorithm but makes it less efficient.

\subsection{Calculating the event free energy}

Once we have a trajectory, it is necessary to compute the free energy difference
between its beginning and its end.

Consider the diagram in Fig.~\ref{fig:tube}. This shows schematically a few nearby activation
paths between different basin boundaries (which are $(N-1)$-dimensional
hyper-surfaces). Points within area $dS_1$ in the figure move to points within
area $dS_2$. If an ensemble with the microcanonical distribution is considered,
the density of flux of the trajectories through a hypersurface is the same for
any hypersurface at all points having the same potential energy. Then the ratio
of the rate of the direct transition (from $dS_1$ to $dS_2$) to the rate of the
inverse transition (from $dS_2$ to $dS_1$) is equal to the ratio of the areas,
$dS_1/dS_2$. If we want the microcanonical distribution to be preserved, this
ratio should equal 1. In general, however, it is not unity and we need to add an
additional acceptance/rejection step, for instance, accepting a particular
activation transition with a Metropolis-like probability $P_{\rm acc}={\rm
  min}(dS_2/dS_1,1)$ for the transition from $dS_1$ to $dS_2$ in the figure.

The activation transition can be considered as a transformation between points
on different basin boundaries (for instance, point $\x_0$ is transformed into
point $\x_p$, area $dS_1$ in Fig.~\ref{fig:tube} is transformed into area
$dS_2$). The ratio $J=dS_2/dS_1$ corresponds therefore to the {\it Jacobian} of
this transformation and it is all we need to ensure detailed balance.

\subsubsection{The Jacobian of the activation transformation - the boundary
  factor}

\begin{figure}
\includegraphics[width=8.5cm]{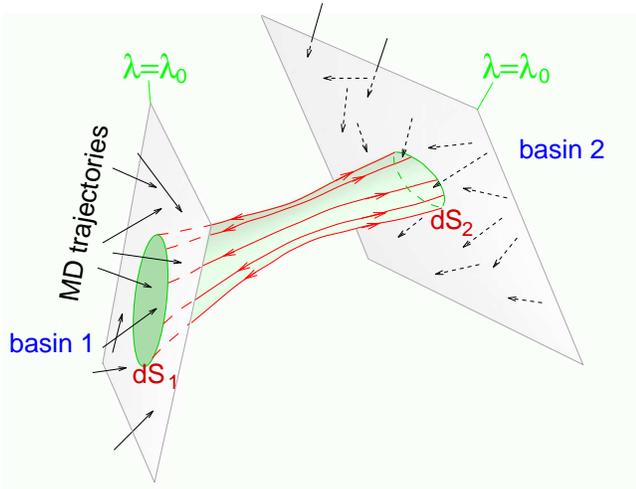}
\caption{ Sketch of tube connecting two basins (see text).}
\label{fig:tube}
\end{figure}

Imagine again a tube formed by nearby activation trajectories
(Fig.~\ref{fig:tube}). If the trajectories forming the tube start within area
$dS_1$ on the first basin boundary, then the cross-section at the beginning of the
tube is $dS'_1=dS_1\cos \alpha_1$, where $\alpha_1$ is the angle between the
normal to the initial basin boundary and the activation trajectory at its start.
Likewise, the cross-section at the end of the tube is $dS'_2=dS_2\cos \alpha_2$,
where $\alpha_2$ is the angle between the normal to the final basin boundary and
the activation trajectory at its end. Including all contributions, the Jacobian
can then be written as
\begin{equation}
J=J_{\rm b} J_{\rm xs},
\end{equation}
where
\begin{eqnarray}
J_{\rm b}&=&{\cos\alpha_1\over \cos\alpha_2},\\
J_{\rm xs}&=&{dS'_2\over dS'_1}.
\end{eqnarray}
We call $J_{\rm b}$ the {\it boundary factor} and $J_{\rm xs}$
the {\it cross-section factor}.

We start by calculating the boundary part $J_{\rm b}$.  First, we note that 
Eq.~(\ref{eq:follow}) for the activation trajectory is the discretized
version of
\begin{equation}
{d\x\over d\tau}=\h(\x)+c(\x)\f(\x).
\label{eq:cont}
\end{equation}
This allows us to get an estimate of $c$, the factor preserving the total energy
during the activation.

We write the change in potential energy as
\begin{equation}
{dU\over d\tau}={d\x\over d\tau}\cdot \nabla U=-{d\x\over d\tau}
\cdot\vec{F}=-\fpa-c\ff,
\end{equation}
where $\fpa=(\f\cdot\h)$.  Since we want to keep $U$ constant, ${dU\over
  d\tau}=0$ and
\begin{equation}
c=-{\fpa\over \ff}.\label{eq:c}
\end{equation}
$\ff$ is a sum over $N$ components of the force and thus scales as ${\cal
  O}(N)$, the system size, since all modes are roughly equally excited.  For its part,
$\fpa$ represents just one component along the activated direction and does not
grow with system size. These observations imply therefore that $c$ scales as
${\cal O}(1/N)$.

Next, we will show that the eigenvector $\h$ is nearly tangent to the trajectory.
Using Eq.~(\ref{eq:cont}), the angle $\beta$ between the activation trajectory
and the eigenvector $\h$ is given by
\begin{equation}
\cos\beta=\frac{(\h\cdot{dx\over d\tau})}{|{dx\over d\tau}|}=
\frac{1+c\fpa}{\sqrt{(1+c\fpa)^2+c^2\cdot \fpp}}=\sqrt{\frac{\fpp}{\ff}}.
\end{equation}
Since, as discussed above, $\fpa^2\ll \ff$ for big systems, $c\fpa,c^2\fpp\ll
1$, and $\cos\beta$ is nearly 1. Thus we can replace the direction of the trajectory
with the direction of $\h$ when calculating angles $\alpha_1$ and $\alpha_2$.

Note that since the basin boundary is by definition the constant-eigenvalue
surface, the normal to it is parallel to $\nabla\lambda$, where $\lambda$ is the
lowest eigenvalue. Then
\begin{equation}
\cos\alpha_{1,2}={\h\cdot\nabla\lambda\over |\nabla\lambda|},
\label{eq:ratio_cos}
\end{equation}
where all quantities are evaluated at the beginning of the activation trajectory
for $\alpha_1$ and at its end for $\alpha_2$.  Thus in order to calculate
$\alpha_1$ and $\alpha_2$, we need a way to find $\nabla\lambda$
numerically. The most efficient method is as follows.  By definition, at point
$\x$ in the configuration space,
\begin{equation}
\hat{H}(\x)\h(\x)=\lambda(\x)\h(\x),
\end{equation}
where $\hat{H}(\x)$ is the Hessian operator at point $\x$. Considering
$\x=(x_1,x_2,\ldots,x_N)$ as a set of $N$ parameters $\{x_i\}$, we can apply the
Hellmann-Feynman theorem and find, in vector form and with Einstein's summation
convention:
\begin{equation}
\nabla\lambda=
\h{\nabla \hat{H}}
\h\equiv {\partial H_{jk}\over\partial x_i}h_j h_k\vec{e}_i = -{\partial^2
\vec{F}\over \partial x_j \partial x_k}h_j h_k,
\end{equation}
where $\vec{e}_i$ are unit vectors along the coordinate axes.

This expression is simply the second derivative of $\vec{F}$
along the direction of the eigenvector $\h$ with the negative sign, i.e.,
\begin{equation}
\nabla\lambda(\x)=\lim_{\delta\to 0} {2\vec{F}(\x)-
\vec{F}(\x+\delta\cdot\h)-\vec{F}(\x-\delta\cdot\h)\over
\delta^2}.\label{eq:jb}
\end{equation}
It can be used directly for numerical evaluation. We simply need to compute the
force $\vec{F}$ at three nearby points along the direction of $\h$ for each boundary 
in order to obtain an accurate evaluation of the boundary factor $J_{\rm b}$.

\subsubsection{Analysis of the cross-section Jacobian}

The second factor in the total Jacobian is the cross-section factor $J_{\rm
  xs}$. To evaluate it, we need to see how the cross-section of an
infinitesimally narrow tube formed by activation trajectories changes between
the two basin boundaries. 
Describing the evolution in the configuration space as a function of $\tau$ by
the equation
\begin{equation}
{d\x\over d\tau}=\vec{f}(\x),
\end{equation}
then, as $\tau$ is incremented by $d\tau$, point $\x$ transforms into
$\x+\vec{f}(\x) d\tau$.  The Jacobian of that transformation is given by the
determinant of the matrix $A_{ij}=\delta_{ij}+{\partial f_i\over \partial
  x_j}d\tau$, which is $1+\sum_i{\partial f_i\over \partial x_i}d\tau+{\cal
  O}(d\tau^2)=1+\dv \vec{f}d\tau+ {\cal O}(d\tau^2)$.  For an infinitesimal
volume $\delta V(\x)$ around point $\x$, the rate of change simply becomes
\begin{equation}
{d\over d\tau}\delta V(\x)=\dv \vec{f}(\x)\cdot\delta V(\x).
\end{equation}
Which can be solved formally:
\begin{equation}
\delta V(\tau)=\delta V(0)\exp\left[\int_0^\tau \dv \vec{f}(\x(\tau'))
d\tau'\right].
\end{equation}

Going back to the continuous version of our evolution equation,
Eq.~(\ref{eq:cont}), we note that $|c\f|=\fpa/|\f|\ll 1$.  Therefore, the speed
along the activation trajectory is nearly constant (equal to one). Thus the
infinitesimal volume $\delta V$ will only change its size and shape in the
transverse directions, but will not shrink or expand in the longitudinal
direction. Then the tube cross-section ratio between any two points on the
trajectory is the same as the volume ratio between the same two points. The
logarithm of the cross-section contribution to the Jacobian is then
\begin{eqnarray}
\ln J_{\rm xs}&=&\ln {\delta V(\tau)\over \delta V(0)}=\int_0^\tau \dv
\vec{f}
(\x(\tau')) d\tau' \label{eq:jxs}\\
&=&\int_0^\tau \dv \h\left(\x(\tau')\right)
+\dv \left[c\left(\x(\tau')\right)\f\left(\x(\tau')\right)\right]
d\tau'.\nonumber
\end{eqnarray}
Equation (\ref{eq:jxs}), together with Eq.~(\ref{eq:jb}) for the boundary
factor, can be used in principle to calculate the activation Jacobian.  However,
straightforward evaluation of the divergences entering Eq.~(\ref{eq:jxs}) by
calculating numerically the derivatives of $\h$ and $\f$ along $N$ orthogonal
directions for many points on the trajectory is computationally very costly and
any usable method will require further careful analysis and making certain
reasonable approximations, as discussed below.

The logarithm of the cross-section factor in the Jacobian is an integral along
the activation trajectory:
\begin{equation}
\ln J_{\rm xs}=\int_0^{\tau} j(\x(\tau')) d\tau',
\end{equation}
where
\begin{equation}
j(\x)=\dv\h+\dv(c\f)=\dv\h+c\,\dv\f+\f\cdot\nabla c.\label{eq:j}
\end{equation}
Compare now the second and the third terms in Eq.~(\ref{eq:j}) 
to show that the third term can be neglected. In the second
term, $\dv\f$ is the trace of the Hessian $H$ taken with the negative sign and
is therefore ${\cal O}(N)$. Since $c$ is ${\cal O}(1/N)$, the second term in
Eq.~(\ref{eq:j}) is ${\cal O}(1)$. Now consider the third term. Using
Eq.~(\ref{eq:c}),
\begin{eqnarray}
\f\cdot\nabla c&=&-\f\cdot\nabla\left(\frac{\fpa}{\ff}\right)\nonumber\\
&=&{\fpa\over\ff}\cdot\frac{\f\cdot\nabla\ff}{\ff}-\frac{\f\cdot\nabla\fpa}{\ff}.
\label{eq:3rd}
\end{eqnarray}
If we use the coordinate system in which axes are parallel to the eigenvectors
of the Hessian at point $\x$ (in particular, the zeroth axis is parallel to
$\h$), then $\partial F_i/\partial x_j=-\lambda_i\delta_{ij}$, where $\lambda_i$
is the $i$th eigenvalue of the Hessian. Then the first term in
Eq.~(\ref{eq:3rd}) is
\begin{equation}
\frac{\fpa}{\ff}\cdot\frac{\f\cdot\nabla\ff}{\ff}=2c\frac{\sum_{i=0}^{N-1}
F_i^2
\lambda_i}{\ff},
\end{equation}
which is ${\cal O}(1/N)$ (given that all $\lambda$'s are ${\cal O}(1)$, $c$ is ${\cal O}(1/N)$ and
$\sum_{i=0}^{N-1} F_i^2=\ff$) and is thus negligible compared to
the second term of Eq.~(\ref{eq:j}). In the second term of Eq.~(\ref{eq:3rd}),
\begin{eqnarray}
\frac{\f\cdot\nabla\fpa}{\ff}&=&\frac{\f\cdot\nabla\left(\sum_{i=0}^{N-1}
F_i h_i\right)}{\ff}\nonumber\\
&=&\frac{\f\cdot\sum_{i,j=0}^{N-1}((\partial F_i/\partial x_j)
h_i\vec{e}_j)}{\ff}
\nonumber\\
& &+\frac{\f\cdot\sum_{i,j=0}^{N-1} (F_i(\partial h_i/\partial
x_j)\vec{e}_j)}{\ff}
\nonumber\\
&=&\frac{-F_0\lambda_0}{\ff}+\frac{\sum_{i,j=0}^{N-1}{\partial
h_i\over\partial x_j}
F_i F_j}{\ff}.
\end{eqnarray}
In the last expression, the first term is clearly ${\cal O}(1/N)$ and thus
negligible; the second term is also negligible (this will be so even
under a completely unrealistic assumption that all of $\partial h_i/\partial
x_j$ are ${\cal O}(1)$, since $F_i$ are of random signs). Thus the third term in
Eq.~(\ref{eq:j}) can always be neglected for big enough $N$ and we end up with
\begin{equation}
j=\dv\h+c\,\dv\f.\label{eq:jfin}
\end{equation}

In Appendix A, we will discuss the physical meaning of the second term
in  Eq.~(\ref{eq:jfin}), using the harmonic approximation.

\subsection{Implementation of the POP-ART algorithm}

The actual implementation of POP-ART, as used to obtain the results presented in
the next section, incorporates the following steps:
\begin{enumerate}

\item We start with MD at finite temperature and first equilibrate by rescaling
  the velocities. We use a 1 fs step and compute the lowest eigenvalue every 10
  steps. After we have crossed the basin boundary defined by the threshold, 
  we retrace our MD path and identify the crossing time with an accuracy of 1 fs.

\item We then apply Eq. (\ref{eq:follow}) and generate the event from one basin
  to another, stopping at the threshold and saving configurations along the way.
  We take $\Delta \tau =0.01\; $\AA.

\item From the first and final configurations of the activation path, we compute
  the boundary factor, $J_{\rm b}$, using Eqs. (\ref{eq:ratio_cos}) and (\ref{eq:jb}).

\item We then evaluate the cross-section Jacobian $J_{\rm xs}$ by integrating $j$, as defined
  by Eq. (\ref{eq:jfin}), over the pathway. We now have the full free energy
  difference between the entry and exit points.

  Straightforward evaluation of the divergences entering Eq.~(\ref{eq:jfin})
  by calculating numerically the derivatives of $\h$ and $\f$ along $N$
  orthogonal directions for many points on the trajectory is computationally
  very demanding.  It is possible, however, to lower this cost significantly
  while keeping a reasonable accuracy. First, we use a 15-iteration Lanczos scheme
  which allows us to obtain the eigenvector $\h$ within ${\cal O}(N)$.  The
  divergence of the eigenvector can also be obtained with ${\cal O}(N)$,
  admittedly with a much larger prefactor, provided the potential is
  sufficiently short-ranged.  For two atomic coordinates $i$ and $j$ belonging
  to atoms which are well outside of each other's interaction range, $\partial
  h_i/\partial x_j\approx 0$. We can exploit this property, to obtain
  $\dv\h=\sum_{i=0}^{N-1}\partial h_i/\partial x_i$ with less than ${\cal O}(N)$
  force evaluations. For example, two terms in this summation can be obtained
  with one eigenvector computation: $\partial h_i/\partial x_i + \partial
  h_j/\partial x_j \approx \sum_{m=\{i,j\}}
  (h_m(\vec{x}+\Delta\vec{e}_i+\Delta\vec{e}_j)-h_m(\vec{x}))/\Delta$. This
  trick can easily be extended as long as the added coordinates are sufficiently
  far apart. For the two systems studied here, the unit cell is divided into 25
  groups of 40 non-interacting atoms each. The total cost of evaluating $\dv\h$
  adds up to 75 Lanczos recursions with only 5 iterations each.
  
\item The previous two steps provide $J_{\rm b}$ and $J_{\rm xs}$
  and thus the free energy difference between the
  first and the last state on the activation trajectory, which is then used in a
  Metropolis accept/reject move. If the event is rejected, the component of the
  velocity normal to the basin boundary is reversed, and MD is continued.
  If the event is accepted, we continue MD in
  the new basin, using the initial velocity (reversing the component of the velocity
  normal to the basin boundary, if necessary). As mentioned above, we run 10 steps to bring
  the configuration away from the border.

\item Once the threshold is reached again, repeat steps 2-6.

\end{enumerate}

\section{Simulation results}

To verify the thermodynamical correctness of the POP-ART method and to
investigate its computational efficiency, we have studied the diffusion
by interstitials and vacancies in a silicon crystal, described by the
Stillinger-Weber potential~\cite{stillinger85} in systems of respectively
1001 and 999 atoms, and a cubic simulation cell of $27.136\; $\AA$^3$.

\subsection{Interstitial diffusion}

First, we look at the interstitial in a 1001-atom cell of Si.  In principle,
many interstitial sites are possible in this system, but only three of them are
significantly populated: the lowest-energy configuration and two configurations with almost the same
energy (within 0.01 eV), about 0.75 eV above the first one~\cite{schober89}. We
are interested in computing the probability of being in one of the higher-energy
states.

If the population of the higher-energy states were determined by the
energy difference alone, this would amount to a population of the
higher-energy states of only 0.07\% at 1200 K. However, there are
degeneracies and the potential wells of the higher-energy states
are much flatter than that for the low-energy state, leading to a
noticeable entropy difference. Because of this difficulty, we extract
the thermodynamical equilibrium between these two states with MD. This
forces us to perform the tests at a relatively high temperature. Here,
we report results for 1000 K and 1200 K.

These conditions are not ideal for POP-ART since at such high temperatures
the jumps between the minima are rather frequent and 
straightforward MD is quite efficient. But given the significance of
both energetic and entropic contributions, as well as some anharmonic effects
present at such high temperatures, it is a very good test for the {\it accuracy}
(rather than efficiency) of POP-ART.

At 1000 K and 1200 K, the system can spend a non-negligible
amount of time outside the basins, i.e., in the saddle region (which would not
be the case in systems more appropriate for POP-ART), leading to a different value
of the ratio of time spent in the upper vs. the lower-energy
states. We therefore need to distinguish carefully between the probability of
being in the {\it attraction region} of a given minimum and the time spent in a
particular {\it basin} (understood as defined in this paper) as a fraction of
the total time spent within all basins. It is the latter quantity that we use to
compare to the POP-ART result.

At 1000 and 1200 K, the MD result for this quantity is, respectively, $1.6\pm 0.1$\% and $3.6\pm 0.1$\%, determined as an
average over 25 runs, each of which lasted 10 ns and an assignment to a basin, 
with threshold $\lambda =-2$ and $-1\; $eV/\AA$^2$, done every 100 fs. 

For POP-ART the fraction is obtained as an average over 5 runs, of 10000
iterations of the POP-ART algorithm each. The assignment to a basin is done
every 100 fs, excluding the 10 steps for moving away from the border after each
generated event. At 1000 and 1200 K, we obtain the respective ratios of $1.3\pm
0.3$\% and $3.5\pm 0.3$\%. Clearly POP-ART samples accurately the
thermodynamical weight of the high and low-energy local mimina.

\subsection{Vacancy diffusion}

Having established the accuracy of POP-ART, we now characterize its
efficiency. For this, we consider a vacancy in a 999-atom cell of Si. Vacancy
diffusion is associated with a single activation barrier of $0.43$
eV~\cite{maroudas93}. Assessing the efficiency of POP-ART relative to MD can be
cleanly done in this system, since the speed of phase space exploration is
simply given by the vacancy's diffusion coefficient, which follows the Arrhenius
law.  The comparison with MD is done on the basis of the number of calls to the
force routine, since that takes $\sim$99\% of the computer time.

\begin{figure}

\includegraphics[width=8.5cm]{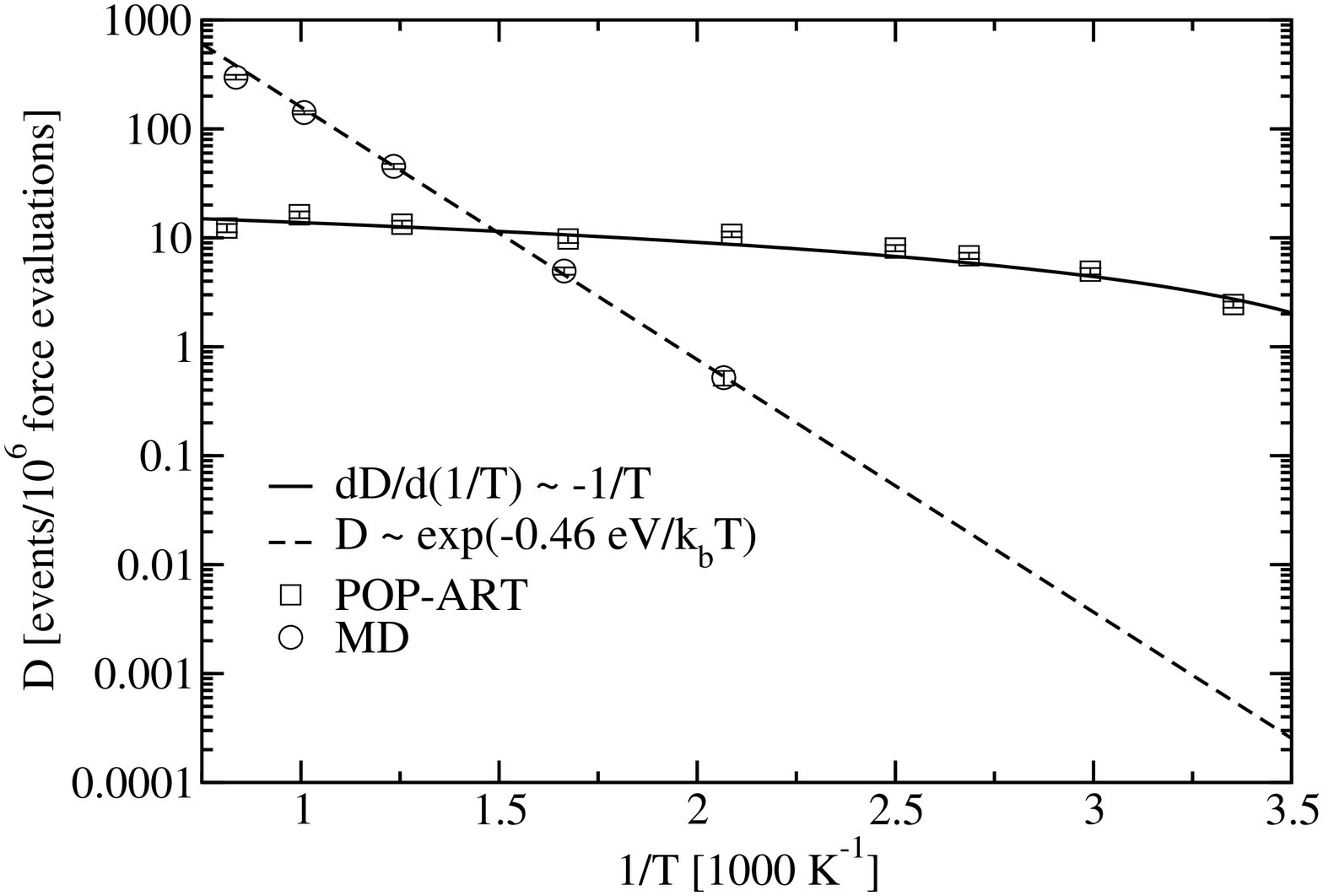}

\caption{Diffusion of a vacancy in Stillinger-Weber silicon (counted as the
  number of jumps per million force evaluations) as a function of temperature
  for molecular dynamics and POP-ART. The dashed line is an Arrhenius fit to the
  MD results with an energy barrier of $0.46$ eV (close to the value of $0.43$ eV
  reported in Ref.~\cite{maroudas93}), and
  the solid line is a $1/T$ fit to the POP-ART results.}
\label{fig:vacancy}
\end{figure}

Figure~\ref{fig:vacancy} shows an Arrhenius plot of the diffusion
rate per million force operations obtained by MD and by POP-ART, as a
function of temperature. The diffusion rate is defined as the vacancy
hopping rate, which is linearly related to the square displacement per
unit time. The value of $\lambda$ used in the POP-ART simulation is selected to optimize
the diffusion rate. The threshold values used in the simulation are $\lambda =-2\; $eV/\AA$^2$
for T $>$ 750 K, $\lambda =-1\;$eV/\AA$^2$ for 600 K $<$ T $<$ 750 K
and $\lambda=0\;$eV/\AA$^2$ for T $<$ 600 K. Clearly, the diffusion
rate per force evaluation obtained with POP-ART does not show activated
behavior and provides a significant boost at temperatures below 700 K,
reaching a factor of more than 4 orders of magnitude at room temperature.
Interestingly, the diffusion rate per force evaluation is not constant
with POP-ART but can be fitted by a $1/T$ curve. The slowing down 
with decreasing temperature is related to the
fact that before POP-ART attempts to find its way to a new basin,
it needs to reach a given curvature threshold. At a very low temperature,
even this will be a very rare event.

\section{Discussion and conclusion}

Efficient sampling of slow systems is one of the main challenges in computational
physics today. For disordered systems such as glasses, for example, standard
thermodynamical methods fail because of the very small phase space occupied by
the relevant configurations. Activated methods, such as ART, overcome these
difficulties by generating physically-possible trajectories through the
conformation space but they do not offer a proper thermodynamical sampling.

The properly-obeying-probability activation-relaxation technique (POP-ART) lifts
these limitations by generating activated trajectories with proper
thermodynamical weighting. Mixing molecular dynamics with activation over
barriers, this algorithm respects detailed balance and samples in a well-defined
thermodynamical ensemble.

To verify the correctness of POP-ART, we sampled the states visited by an
interstitial in {\it c}-Si. In this system, the higher-energy states of the
interstitial are energetically suppressed but entropically favored. Comparing
with MD, we found that POP-ART samples the high-energy states with the proper
probability, demonstrating its correctness. In order to assess the efficiency of
this method, we looked at the vacancy diffusion also in {\it c}-Si. In this
case, POP-ART is found to outperform MD below 700 K, and it is about four orders
of magnitude faster at room temperature.

One of the advantages of POP-ART is that all the information it needs is
local. This makes it possible to apply a number of approximations to increase
further its efficiency without sacrificing the sampling.  POP-ART can also be
extended to reproduce the correct activated dynamics; this is currently examined
and will be reported in a further publication.

\section*{Acknowledgements}

MC is supported in part by the Fonds qu\'eb\'ecois de recherche sur la nature et
les technologies (FQRNT). NM acknowledges support by NSERC (Canada), FQRNT
(Qu\'ebec), the Research Corporation (USA), and the visiting scientist program of the
NWO (the Netherlands).

\section*{Appendix A: The physical meaning of the divergence of the force}

Using the harmonic approximation, we can reveal better the physical meaning of
the second term in Eq.~(\ref{eq:jfin}). First of all, using the expression for
$c$, Eq.~(\ref{eq:c}), we get
\begin{equation}
c\,\dv\f=-\frac{\fpa}{\ff}\dv\f.
\end{equation}

At relatively low temperatures, the system is well described by the harmonic
approximation. The potential around the minimum can then be rewritten as
\begin{equation}
V=V_0+{1\over 2}\sum_{i=0}^{N-1} k_i x_i^2,
\end{equation}
where $x_i$ represent the normal modes, and the force becomes
\begin{equation}
\dv\f\approx-\sum_{i=0}^{N-1}k_i.\label{eq:divf}
\end{equation}
Using the same approximation for $\ff$ and neglecting anharmonicity, we get:
\begin{equation}
\ff\approx\sum_{i=0}^{N-1}k_i^2 x_i^2.\label{eq:f2}
\end{equation}
Since the spectrum should be dense, $x_i^2$ can be replaced with their thermal
averages $\langle x_i^2\rangle=k_BT/k_i$, giving
\begin{equation}
\ff\approx k_BT\sum_{i=0}^{N-1}k_i,
\end{equation}
and
\begin{equation}
c\,\dv\f\approx {\fpa\over k_B T}.\label{eq:fpa}
\end{equation}
After integrating over the whole activation trajectory, we obtain
\begin{equation}
\int c\,\dv\f d\tau'=-{\Delta E_{\parallel}\over k_BT},
\end{equation}
where
\begin{equation}
\Delta E_{\parallel}=-\int\fpa d\tau'.
\end{equation}
Then the contribution of the $c\,\dv\f$ term to the Jacobian $J$ is
\begin{equation}
\exp\left(\int c\,\dv\f d\tau'\right)=\exp\left(-{\Delta E_{\parallel}\over
k_BT}\right),
\end{equation}
a Boltzmann factor. $\Delta E_{\parallel}$ is essentially the energy change that
would have occurred along the activation trajectory, if it were parallel to $\h$
everywhere and the energy-correcting $c\f$ term in Eq.~(\ref{eq:follow}) was not
there.

The transition probability between two minima should contain both energetic and
entropic contributions. Given the above result, it is tempting to associate the
$\dv\h$ term with the entropic and the $c\,\dv\f$ with the energetic contribution;
however, an example considered in Appendix B shows that the reality is more
complex: in fact, $\Delta E_{\parallel}$, defined as above, is
temperature-dependent and the $c\,\dv\f$ term therefore contains 
both energetic and the entropic contributions.

\section*{Appendix B: the Jacobian in a model potential}

To get some insight into the physical meaning of the activation Jacobian and its
different components, consider the following model, defined by the potential:
\begin{equation}
U=U_0(x_0)+{1\over 2}\sum_{i=1}^{N-1} k_i(x_0)x_i^2.\label{eq:pot}
\end{equation}
Here $U_0(x_0)$ is a function with two minima and a maximum between them,
so that
coordinate $x_0$ describes the activated mode (i.e., is the
``reaction coordinate''), and the other $N-1$ degrees of freedom
serve as the ``heat bath''. ``Force constants'' $k_i(x_0)$ are assumed
to remain positive for all $x_0$ of interest (e.g., between the minima).

Our model does not represent the most general
situation. In particular, the eigenmode-following transition path (such as,
e.g., ART would find) is a straight line (coinciding with the 0th axis);
also,
on that line, an eigenvector for a particular mode has the same direction
(parallel to a
coordinate axis) everywhere. However, the model is interesting enough: as
the
frequencies of the bath modes (determined by $k_i(x_0)$) can change along
the
activation path, there are both energetic and entropic contributions to the
probability of being in a particular place along the reaction coordinate.
Indeed, the probability density of having the zeroth coordinate equal to
$x_0$
is
\begin{eqnarray}
p(x_0)&\propto& \int dx_1\ldots dx_{N-1} \exp[-U/k_B T]\nonumber\\
&=&\exp\left[-\frac{U_0(x_0)}{k_B T}\right]\nonumber\\
& &\times\int dx_1\ldots dx_{N-1}
\exp\left[-\sum_{i=1}^{N-1}
\frac{k_i(x_0)x_i^2}{2k_B T}\right]\nonumber\\
&=&\exp\left[-\frac{U_0(x_0)}{k_B T}\right]\prod_{i=1}^{N-1}
\left(\frac{2\pi k_B T}{k_i(x_0)}\right)^{1/2},
\end{eqnarray}
or
\begin{equation}
p(x_0)\propto \exp[-{\cal F}(x_0)/k_B T],
\end{equation}
where the free energy
\begin{equation}
{\cal F}(x_0)=U_0(x_0)-TS(x_0)\label{freeen}
\end{equation}
and the entropy
\begin{equation}
S(x_0)=-{1\over 2}k_B \sum_{i=1}^{N-1} \ln k_i(x_0).\label{entropy}
\end{equation}

We will assume in what follows that $k_i(x_0)$ are linear functions, i.e.,
\begin{equation}
k_i(x_0)=k_i^{(0)}+k_i^{(1)}x_0.
\end{equation}
The matrix elements of the Hessian for the potential given by
Eq.~(\ref{eq:pot}) are
\begin{eqnarray}
H_{00}&=&U''_0(x_0),\nonumber\\
H_{ii}&=&k_i(x_0),\ \ i\ne 0,\\
H_{0i}=H_{i0}&=&k_i^{(1)}x_i,\nonumber
\end{eqnarray}
and the remaining elements are zero. The force components are
\begin{eqnarray}
F_0&=&-U_0'(x_0)-{1\over 2}\sum_{i=1}^{N-1} k_i^{(1)}x_i^2,\nonumber\\
F_i&=&-k_i(x_0)x_i,\ \ i\ne 0.\label{eq:force}
\end{eqnarray}

For a Hessian with only $H_{ii}$ and $H_{0i}=H_{i0}$
non-zero,
\begin{equation}
\h=C\left(1,\frac{H_{01}}{\lambda-H_{11}},\frac{H_{02}}{\lambda-H_{22}},\ldots
\right),
\end{equation}
where $C$ is the normalization constant, and the eigenvalue $\lambda$ is
the
lowest solution of the following equation:
\begin{equation}
\lambda=H_{00}+\sum_{i=1}^{N-1} \frac{H_{0i}^2}{\lambda-H_{ii}}.
\end{equation}

In a real physical system, the entropy change along an activation
trajectory will always remain finite and of the same order of magnitude as the thermal
energy, as the system size increases. In our model, this will
mean that most $k_i^{(1)}$ are small enough. Note that this is essentially
the
same as the assumption of most modes being
nearly harmonic that we have used when approximating the $c\,\dv\f$ term in
the
Jacobian. Specifically,
\begin{equation}
T\Delta S\sim k_BT\sum_{i=1}^{N-1}{k_i^{(1)}\over k_i}\Delta x_0 \sim k_BT,
\end{equation}
or
\begin{equation}
\sum_{i=1}^{N-1}{k_i^{(1)}\over k_i}\Delta x_0\sim 1.\label{k1}
\end{equation}
For simplicity, we will assume in addition that
\begin{eqnarray}
\sum_{i=1}^{N-1} \frac{H_{0i}^2}{H_{00}-H_{ii}}&\ll& H_{00},\\
\sum_{i=1}^{N-1} \frac{H_{0i}^2}{(H_{00}-H_{ii})^2}&\ll& 1.
\end{eqnarray}
This will be the case, in particular, at low enough $T$, when the
magnitudes
of most $x_i$ are small. Under these conditions,
\begin{eqnarray}
\lambda&\approx& H_{00}=U''_0(x_0),\\
C&\approx& 1.
\end{eqnarray}
The constant-eigenvalue surfaces are then orthogonal to the 0th axis and
the
cosine of the angle between the normal to a constant-eigenvalue surface and
$\h$ is nearly 1, so the boundary contribution to the activation Jacobian
can
be neglected. Consider now the cross-section factor. We need to calculate
$j$, as given by Eq.~(\ref{eq:jfin}). First of all,
\begin{equation}
\h=\left(1,\frac{k_1^{(1)}x_1}{U_0''(x_0)-k_1(x_0)},\ldots\right)
\end{equation}
and so
\begin{equation}
\dv\h=\sum_{i=1}^{N-1}\frac{k_i^{(1)}}{U_0''(x_0)-k_i(x_0)}.
\end{equation}
Further, using Eq.~(\ref{eq:force}),
\begin{eqnarray}
\fpa&\equiv&(\f\cdot\h)=-U'_0(x_0)-{1\over 2}\sum_{i=1}^{N-1}
k_i^{(1)}x_i^2
\nonumber\\
&
&-\sum_{i=1}^{N-1}\frac{k_i(x_0)k_i^{(1)}x_i^2}{U_0''(x_0)-k_i(x_0)}\nonumber\\
&=&-U_0'(x_0)-{1\over
2}\sum_{i=1}^{N-1}\frac{U_0''(x_0)+k_i(x_0)}{U_0''(x_0)-
k_i(x_0)}k_i^{(1)}x_i^2.
\end{eqnarray}
We can now use Eq.~(\ref{eq:fpa}) to calculate $c\,\dv\f$; it is, however,
instructive to repeat the considerations, to see how exactly the
appropriate
approximations are made in this particular case. We obtain
\begin{equation}
\dv\f=-U_0''(x_0)-\sum_{i=1}^{N-1}k_i(x_0)\approx
-\sum_{i=1}^{N-1}k_i(x_0),
\end{equation}
(where an approximation similar to Eq.~(\ref{eq:divf}) is obtained by
neglecting the
term associated with the anharmonic activated mode) and
\begin{equation}
\ff=\left(U_0'(x_0)-{1\over 2}\sum_{i=1}^{N-1}k_i^{(1)}x_i^2\right)^2+
\sum_{i=1}^{N-1} k_i^2(x_0)x_i^2.
\end{equation}
In the last expression, bearing in mind the condition (\ref{k1}), the
second
term in the parentheses is of the same order of magnitude as the first one,
and then both of them are negligible compared to the second sum (this is an
approximation similar to the one used in Eq.~(\ref{eq:f2})). We therefore
obtain
\begin{eqnarray}
c\,\dv\f&\approx& -\left[U_0'(x_0)+{1\over
2}\sum_{i=1}^{N-1}\frac{U_0''(x_0)+k_i(x_0)}{U_0''(x_0)-
k_i(x_0)}k_i^{(1)}x_i^2\right]\nonumber\\
& &\times\frac{\sum_{i=1}^{N-1}k_i(x_0)}
{\sum_{i=1}^{N-1} k_i^2(x_0)x_i^2}.\label{cdivF}
\end{eqnarray}
As explained in Appendix A, we can replace $x_i^2$ with their thermal
averages and then
\begin{eqnarray}
c\,\dv\f &\approx& -\Big[U_0'(x_0)\\
& &\left.+{1\over 2}\sum_{i=1}^{N-1}
\frac{U_0''(x_0)+k_i(x_0)}{U_0''(x_0)-k_i(x_0)}\frac{k_i^{(1)}}{k_i(x_0)}
k_BT\right]\Big/k_BT.\nonumber
\end{eqnarray}
Then finally
\begin{equation}
j = -U_0'(x_0)/k_BT-{1\over 2}\sum_{i=1}^{N-1}
{k_i^{(1)}\over k_i},
\end{equation}
and the Jacobian $J$ is
\begin{eqnarray}
J&=&\exp(\smallint j(x_0)dx_0)=\exp\left[-(\Delta U_0-T\Delta
S)/k_BT\right]
\nonumber\\
&=&\exp[-\Delta{\cal F}/k_BT],
\end{eqnarray}
where $\Delta U_0$ is the change in $U_0$, $\Delta S$ is the change in
entropy
$S$ [as given by Eq.~(\ref{entropy})] between the two basin boundaries, and
$\Delta{\cal F }=\Delta U_0-T\Delta S$ --- exactly as expected.
\bibliography{bibliographie}
%







\end{document}